# Deciphering the lattice vibrational behaviors of $CuInP_2S_6$ by angle-resolved polarized Raman scattering

Yiqi Hu (胡一奇)[1], Jun-Jie Zhang (张骏杰)[2,*], Zhou Zhou (周洲)[1], Shun Wang (王顺)[3], Qiankun Li (李乾坤)[1], Yanfei Hou (侯艳妃)[1], Tianhao Ying (应天浩)[1], Lingling Yang (杨玲玲)[1], Jingyao Zhang (张静瑶)[1], Shuzhong Yin (殷树中)[1], Yuyan Weng (翁雨燕)[1], Shuai Dong (董帅)[2], Jianlin Yao (姚建林)[4], Liang Fang (方亮)[1,*], Lu You (游陆)[1,3,*]

[1]School of Physical Science and Technology, Jiangsu Key Laboratory of Frontier Material Physics and Devices, Soochow University, Suzhou, 215006, China

[2]Key Laboratory of Quantum Materials and Devices of Ministry of Education, School of Physics, Southeast University, Nanjing 211189, China

[3]School of Electronic and Information Engineering, Suzhou University of Technology, Changshu 215500, PR China

[4]College of Chemistry, Chemical Engineering & Materials Science, Soochow University, Suzhou, 215006, China

*Corresponding authors:

junjiezhang@seu.edu.cn;

lfang@suda.edu.cn;

lyou@suda.edu.cn

## Abstract

The layered van der Waals (vdW) ferroelectric $CuInP_2S_6$ (CIPS) exhibits unique cation-hopping-driven phenomena that bring about unconventional properties with intriguing mechanisms and hold promises for advanced applications in nanoelectronics. However, an explicit analysis of its lattice dynamics and vibrational symmetries, pivotal for understanding the material's peculiar ferroelectric and ferroionic behaviors, remains incomplete. Here, we employ angle-resolved polarized Raman spectroscopy in concert with first-principles calculations to systematically unravel the anisotropic lattice vibrations of CIPS single crystals. By analyzing the polarization-dependent Raman intensities, we determine the symmetry assignments and Raman tensors of all major vibrational modes, revealing good agreement with theoretical predictions. Furthermore, we demonstrate the utility of Raman spectroscopy as a sensitive and non-invasive probe for structural and ferroelectric order evolution, by examining temperature-driven phase transitions and thickness-dependent polarization suppression in CIPS. Our findings establish a foundational framework for correlating lattice dynamics with functional properties in CIPS and provide a methodological blueprint for studying other vdW ferroelectrics.

## Introduction

The layered van der Waals (vdW) material $CuInP_2S_6$ (CIPS) has garnered significant attention in the past decade due to its room-temperature ferroelectricity, piezoelectric response, and ion-conduction capabilities, which hold great promise for applications in flexible nano-electronics, energy-efficient memory, and neuromorphic computing [1-5]. Belonging to the family of metal thiophosphates, CIPS crystallizes in a lamellar architecture characterized by alternating planes of $Cu^+$ and $In^{3+}$ cations interleaved with $P_2S_6^{4-}$ anionic motifs [6]. The tendency of monovalent $Cu^+$ ion for lower coordination drives its off-centered displacement from the center of S octahedron towards top/bottom S triangular planes, resulting in switchable electric polarization. However, the displacive instability of $Cu^+$ further enables its hopping into or even across the vdW gap [7]. This unique feature underpins the abnormal physical properties of CIPS, such as negative longitudinal piezoelectricity [8, 9], quadruple polarization state [10] and potential ferroionic behavior [11].

Raman spectroscopy serves as a vital tool for unravelling the lattice dynamics of CIPS, which is crucial for gaining in-depth understanding of its intriguing cation hopping behaviors. The technique probes vibrational modes associated with the $P_2S_6^{4-}$ units, cation ordering, and interlayer interactions, which are directly tied to the material's ferroelectric phase transitions, domain dynamics, and response to external stimuli (e.g., temperature, electric fields, or strain). However, previous Raman spectroscopic studies have focused mainly on the mode assignments of the vibrational peaks and their evolution across the phase transition point [12-14]. A thorough analysis of the tensorial properties of the vibrational modes is still missing in the literature, which hinders the correct identification of the phonon symmetries and the understanding of related properties' anisotropies.

To this end, we performed a compressive study of the lattice vibrational properties of CIPS single crystal using orientation and temperature dependent angle-resolved polarized Raman scattering, from which the symmetry of the vibrational modes and their Raman tensors can be determined. The experimental results are generally in line with the theoretical calculations, except for some discrepancies which can be successfully explained by the structural disorders originated from the hopping motion of Cu ions. The accurate assignments and direct visualizations of all main vibrational modes in CIPS lay a solid foundation for understanding structure and property evolution as a function of various parameters. To exemplify, Raman spectroscopy was used as a probe to understand the temperature-driven phase transition and thickness-dependent ferroelectric order in CIPS, which are elaborated below.

## Results

The experimental setup of Raman scattering used in this work is illustrated in **Figure**

**1a**. Since CIPS possesses a monoclinic crystal structure, an orthogonal crystallographic coordinate, ***a (X), b (Y),*** $c^*$ ***(Z)***, is used in the study to reduce complexity, where $c^*$ denotes the reciprocal axis perpendicular to ***a*** and ***b*** axis. Next, the angle-resolved Raman scattering measurements were carried out on different crystallographic planes, namely, $bc^*$ $(30\bar{1})$, $ac^*$ (001), and ***ab*** (001) plane of a CIPS bulk single crystal with well-defined facets in back-scattering mode. The incidence of light is vertical to the crystal plane with its electric field lying within the plane, probing the tensorial polarizability change during vibration[15]. The contour maps of angle-dependent polarized Raman intensities under parallel polarization configuration are displayed in **Figure 1(e-g)** for different crystallographic planes (see **Figure S1** for separate spectra), while **Figure 1(b-d)** show the corresponding plane views of the crystal structure in each case. At first glance, contour maps of $bc^*$ and $ac^*$ planes exhibit strong angular anisotropy, unlike the ***ab*** plane, which shows negligible angular dependence. This observation is fully expected from the crystal structures shown above, as the polarizability changes along two intralayer axes, ***a*** and ***b***, should be very different from that along the interlayer $c^*$ axis. The negligible intralayer anisotropy may appear inconsistent with the monoclinic lattice symmetry, but can be understood from a comprehensive analysis of its crystal structure. For *monolayer* CIPS, it possesses isotropic three-fold rotational (C3) symmetry in the layer plane, which, however, is broken by the interlayer stacking pattern. In particular, the adjacent layer first has its metal sites rotated for 180° around In site as the center, and stacks atop with a horizontal shift of 1/3***a***. Such stacking operation destroys the in-plane C3 symmetry, yet resulting in in-plane anisotropic lattice featured by aligned In atomic columns along ***a*** axis (**Figure 1d**) and a glide plane (**m**) perpendicular to ***b*** axis (**Figure 1b**). Since this in-plane structural anisotropy arises from the interlayer interaction (vdW force), it only weakly affects the lattice vibration, leading to subtle but distinguishable differences between ***a*** and ***b*** axes, as reported previously [16].

Having established an intuitive correlation between the crystal structure and Raman polarimetric spectra, we then proceed with an in-depth analysis of each vibrational mode by focusing on the anisotropic ***bc****-plane map, which may reveal more information on the vibrational symmetry. First-principles density functional theory (DFT) calculations were employed to determine the displacement patterns, symmetries and tensor elements of each mode (**Table S1**). The primitive cell of *Cc* phase of CIPS contains two formula units (20 atoms), yielding 60 phonon modes (3 acoustic and 57 optical). According to group theory analysis [17], the irreducible representation of the optical phonon modes is $\Gamma_{optic} = 28\ A' + 29\ A''$, wherein A' and A'' are symmetric and anti-symmetric vibration modes respectively with regard to the glide plane (**m**), as illustrated in **Figure 1b**. Based on the calculated Raman tensors, the angle-dependent Raman spectra of ***bc**** plane can be derived and compared with the experimental one, as shown in **Figure 2(a, b)**. The general features of the experimental and simulated spectra match reasonably well with each other as indicated by the labelled peaks ($P_1$-$P_{15}$), which enables a tentative assignment of each peak to corresponding vibrational mode (**Table S1**). However, experimental high-frequency peaks ($P_{12}$-$P_{15}$) exhibit significant blueshifts compared to theoretical predictions. These peaks mainly arise from the stretching modes of P-P or P-S bonds in the $P_2S_6^{4-}$ unit, whose frequencies are highly sensitive to bond lengths, and the calculated bond lengths are larger than those observed experimentally. A possible explanation for this disparity is that the experimental spectra were measured at room temperature, when Cu ions exhibited significant thermal hopping motions that would compress the bond lengths of $P_2S_6^{4-}$ unit. The theoretical calculations, however, were conducted at 0 K with harmonic approximation and completely no $Cu^+$ hopping motions. The discrepancies in peak intensities can also be explained by the same argument. For example, $P_1$ mode is identified as out-of-plane translation of Cu ions, whose intensity is strongly reduced in actual result due to the smearing nature of the Cu site occupancy. In contrast, $P_9$ mode is featured by in-plane twisting of the $PS_3$ unit and concurrent stretching of Cu-S bonds. It appears that the Cu hopping motion will greatly enhance its Raman activity, thus resulting in much stronger peak intensity in experiment than in calculation.

Next, angle-dependent peak intensity of each mode was extracted from the spectra and plotted in polar coordinates. Raman peak intensity is governed by the projection of the Raman tensor onto the experimental polarization vectors of the light field, as can be expressed as follows,

$$I \propto \left|\vec{e_i} \cdot \boldsymbol{R} \cdot \vec{e_s}^{T}\right|^2, \tag{1}$$

where $\vec{e_i}$ and $\vec{e_s}$ represent the electric field vectors of the incident and scattered light, respectively, and **R** denotes the Raman tensor of specific vibrational modes. In parallel-polarized setup ($\vec{e_i} \parallel \vec{e_s}$), $\vec{e_i} = \vec{e_s} = (0, \cos\theta, \sin\theta)$, while in cross-polarized setup ($\vec{e_i} \perp \vec{e_s}$), $\vec{e_i} = (0, \cos\theta, \sin\theta)$, $\vec{e_s} = (0, -\sin\theta, \cos\theta)$, in which $\theta$ stands for the azimuthal angle between light field vector and ***b*** axis. For a given vibrational mode, the Raman tensor $R$ is a 3×3 matrix representing the change in molecular polarizability during the vibration. Because the lattice structure of CIPS belongs to $m$ ($C_s$) point group, symmetry constraints dictate that the Raman tensors adopt the following forms,

$$\boldsymbol{R}\,(A') = \begin{pmatrix} ae^{i\Phi_a} & 0 & de^{i\Phi_d} \\ 0 & be^{i\Phi_b} & 0 \\ de^{i\Phi_d} & 0 & ce^{i\Phi_c} \end{pmatrix},\ \boldsymbol{R}\,(A'') = \begin{pmatrix} 0 & fe^{i\Phi_f} & 0 \\ fe^{i\Phi_f} & 0 & ge^{i\Phi_g} \\ 0 & ge^{i\Phi_g} & 0 \end{pmatrix},$$

where $a$, $b$, $c$, $d$, $e$, and $f$ are non-zero tensor elements. Additionally, an $e^{i\Phi}$ term ($\Phi$ is the corresponding phase) is added in the Raman tensor to account for its complex nature [18, 19]. Experimentally derived Raman tensor elements may differ from intrinsic Raman tensors due to extrinsic effects such as birefringence, linear dichroism and multilayer interference [20]. In this work, the interference effect and linear dichroism can be ignored due to bulk sample and below-bandgap excitation. As a result, the Raman tensor elements obtain here are effective ones determined by combined intrinsic Raman tensors and birefringence effect. The angle-dependent Raman intensities in the parallel-polarized setup can be calculated using the following formulae,

$$I\left(A'\right)_{\parallel} \propto b^2\cos^4\theta + c^2\sin^4\theta + 2bc\cos\Phi_{bc}\sin^2\theta\cos^2\theta, \tag{2}$$

$$I\left(A''\right)_{\parallel} \propto (g\sin 2\theta)^2, \tag{3}$$

where $\Phi_{bc}$ is $\Phi_c - \Phi_b$. Apparently, Raman intensities of A'-symmetry modes should have a period of π, while those of A''-symmetry modes should show π/2 angular dependence. Experimental intensity data of all major Raman peaks were then fitted to above equations as summarized in **Table S2**, among which all peaks except for $P_7$ are attributed to A' symmetry modes. The results of some representative peaks are displayed in **Figure 2(c-h)**. The ratios of diagonal Raman tensor elements (b/c), that represent the anisotropic polarizability changes along the principal axes (***b*** and ***c***$^*$), can be extracted from the theoretical fits. The experimental and DFT-calculated results generally agree with each other, further corroborating the accuracy of previous mode assignments. The only large discrepancy occurs for $P_{11}$, which accounts for the in-plane stretching of Cu-S bonds. DFT calculation overestimates the anisotropy of this mode due to the omission of out-of-plane Cu hopping motion. Besides, Raman polarimetry of each mode was also performed on ***ac***$^*$ plane, which gives similar results as those of ***bc***$^*$ plane (**Figure S2**). This is fully expected because the anisotropy between ***a*** and ***b*** axes are quite small.

The decisive mode assignments of the major Raman peaks and direct visualization of their displacement patterns greatly facilitate our analyses of the phase transition using Raman spectra as the structural probe. Shown in **Figure 3a** is the temperature-dependent Raman spectra recorded under $Z(XX)\bar{Z}$ configuration, namely, with light incidence/scattering along ***c***$^*$ (Z) axis and its electric field along ***a*** (X) axis. The frequency (in wavenumbers) and line width (FWHM) for each peak can be derived from peak deconvolution fitting, and then plotted as a function of temperature (**Figure 3b, 3c**). It should be noted that the measurement was conducted on cooling branch of the thermal cycle. Therefore, the phase transition temperature (~ 300 K) is slightly lower than those commonly reported (~310 K) [21-23]. $P_{12}$ peak represents the stretching of P-P bonds, a signature internal mode of the $P_2S_6$ unit in metal thiophosphates, which can thus serve as a structural reference for other modes. Across all temperature-dependent curves, the linewidth increases monotonically with temperature, as expected due to progressive structural disorder at higher temperatures, particularly upon activation of $Cu^+$ hopping. Before ferroelectric-paraelectric (FE-PE) phase transition (130 – 250 K), the vibrational frequencies of all Raman peaks show almost linear redshift with increased temperature. This is a typical consequence of thermal expansion, resulting in the softening of the chemical bonds due to their anharmonicity nature [24].

The thermal coefficients of different Raman peaks in this temperature range are typically around -0.01 to -0.02 $cm^{-1}/K$ (**Table S3** and **Figure S3**), consistent with previous report [25]. Upon entering the phase transition region, the redshift rates of most peaks increase by more than one order of magnitude, in accordance with the sudden rise of the lattice parameters along three axes owing to the intralayer Cu hopping between top and bottom sites [7]. As a consequence, in-plane bond stretching modes (e. g. $P_9$, $P_{13}$, $P_{14}$) show more prominent redshifts because the intralayer structure becomes less compact with softened oscillator spring. In stark contrast, out-of-plane layer expansion modes, such as $P_7$ and $P_8$, are featured by apparent blueshifts during phase transition. This behavior could be attributed to the Cu hopping into vdW gap, promoting the interlayer interaction with the S atoms in adjacent layer. Consequently, the spring constant of the interlayer vdW interaction strengthens, resulting in the increase of vibrational frequency. The abnormal increase of the frequency of $P_{11}$ (in-plane asymmetric stretching of Cu-S bonds) can be explained by the contraction of Cu-S bond length due to reduced Cu occupancy [7]. In terms of peak intensity, since it is highly sensitive to various factors during measurement, we analyzed the intensity ratio of each peak against $P_{12}$, which is relatively unaffected during phase transition (**Figure S4**). Some peaks ($P_6$, $P_7$, $P_8$, $P_{11}$) show dramatic intensity decrease in paraelectric phase due to the restoration of inversion symmetry. Furthermore, we have also measured the temperature dependent Raman spectra with light field along $\boldsymbol{c}^*$ (Z) axis in $X(ZZ)\bar{X}$ configuration. However, the poor spectral intensity only allows us to extract the thermal evolution of some major peaks, whose behaviors are generally consistent with the results of the $Z(XX)\bar{Z}$ case (**Figure S5**).

In addition to thermally-driven phase transition, structural change upon thickness reduction was also speculated for CIPS, which leads to the loss of in-plane polarization [26]. To verify this size effect, a combined study using Raman spectroscopy and piezoresponse force microscopy (PFM) was carried out on exfoliated CIPS flakes with varying thickness (6 – 100 nm, **Figure S6**). Gold-coated Si substrate was used to enhance the Raman scattering signal through surface plasmonic resonance effect [27].

For all the samples measured, we observed similar spectral patterns with fingerprint peaks consistent with those of the bulk (**Figure 4a**), whereas the trigonal $P\bar{3}1c$ phase is characterized by a completely different Raman spectrum as cross-checked by both our high-pressure Raman measurement [28] and DFT calculation (**Figure 4c**). This finding suggests the CIPS thin flakes retain the monoclinic *Cc* structure down to very small thickness. Nevertheless, the thickness-dependent Raman spectra reveal additional information on the ferroelectric order parameter of CIPS thin flakes. Particularly, the intensity ratios of $P_6/P_{12}$ and $P_{11}/P_{12}$ decrease drastically for thickness below 50 nm. These two modes show prominent intensity reduction during FE-PE phase transition, thus serving as vibrational indicators for the magnitude of ferroelectric polarization (degree of Cu ordering) [29]. In good agreement, piezoresponse amplitude, which is proportional to the polarization magnitude, exhibits similar thickness dependence for different CIPS flakes due to depolarization effect. Piezoresponse and opposing domains vanish in flakes below 10 nm, a critical thickness greater than previously reported [30]. This discrepancy may arise from the rough gold film (roughness ≈ 2 nm) used here, which inadequately screens polarization charges, enhancing the depolarization field.

The temperature- and thickness-dependent case studies presented above clearly demonstrate Raman spectroscopy's utility as a sensitive, accurate, and non-invasive probe for tracking structural and polar order parameter evolution in CIPS. This analytical approach can be readily extended to other applications, including in situ strain monitoring, domain switching studies, ionic conduction characterization, and operando device measurements [31-34]. However, such applications fundamentally require unambiguous assignment of Raman peaks to specific vibrational modes and a complete understanding of their displacement patterns. Our work successfully establishes a comprehensive framework for analyzing lattice vibration characteristics in prototypical vdW ferroelectrics through the combined use of angle-resolved polarized Raman spectroscopy and DFT calculations, a methodology directly transferable to other material systems.

## Methods

### Materials synthesis and sample preparation

Phase-pure $CuInP_2S_6$ single crystals were synthesized by chemical vapor transport (CVT) method as reported previously [30]. For surface-enhanced Raman measurements, 10-nm-thick gold film was deposited on $SiO_2$/Si substrate, followed by mechanical exfoliation of CIPS thin flakes onto it.

### Raman spectroscopy

Polarized Raman spectra were measured in back-reflection mode by integrated confocal Raman spectrometer (Horiba, Xplora Plus for standard measurements and WITec, Alpha 300RAS for low-wavenumber measurements) with different sample rotation angles using a rotational stage. A 532-nm laser was used for excitation with the polarizer and analyzer fixed in parallel-polarized configuration, and 2000 gr/mm grating was used for the spectrometer. Temperature dependent Raman scattering measurements were conducted using a cryogenic micromanipulator probe station equipped with a heating stage (Instec, HCS621G).

### Piezoresponse force microscopy

Vertical PFM measurements were conducted on a commercial atomic force microscope (Oxford Instruments, MFP-3D Origin+) in ambient condition using Pt/Ir-coated probes (resonant frequency ≈ 70 kHz, spring constant ≈ 2 N/m). Dual AC resonance tracking (DART) method was employed by driving the tip at the vertical contact resonance near 300 kHz with driving voltage $V_{ac}$ = 1 V. The effective piezoresponse of PFM image was obtained through optical lever sensitivity calibration and fitting to simple harmonic oscillator (SHO) model.

### Computational method

Density functional theory (DFT) calculations were carried out using the plane-wave basis set as implemented in the Vienna ab initio Simulation Package (VASP) [35]. The exchange-correlation effects were treated using the generalized gradient approximation in the Perdew–Burke–Ernzerhof functional [36], with an energy cutoff of 500 eV. The Brillouin zone was sampled using a 6 × 6 × 3 Monkhorst–Pack *k*-point mesh. Phonon frequencies were computed using density functional perturbation theory as implemented in VASP. The Raman tensor was obtained from the macroscopic dielectric response [37].

## Supporting Information

Supporting Information includes additional experimental data and figures.

## Acknowledgements

L.Y. and L.F. acknowledge the support by the National Natural Science Foundation of China (No. 12474089 and No. 12574102) and the Priority Academic Program Development (PAPD) of Jiangsu Higher Education Institutions. J.-J. Z. acknowledges support by National Natural Science Foundation of China (Grants No. 12404102), Natural Science Foundation of the Jiangsu Province (Grant No. BK20230806), and Southeast University Interdisciplinary Research Program for Young Scholars (Grant No. 2024FGC1008).

## Conflict of Interest

The authors declare no conflicts of interest.

## Data Availability Statement

The data that support the findings of this paper are available from the corresponding authors upon request.

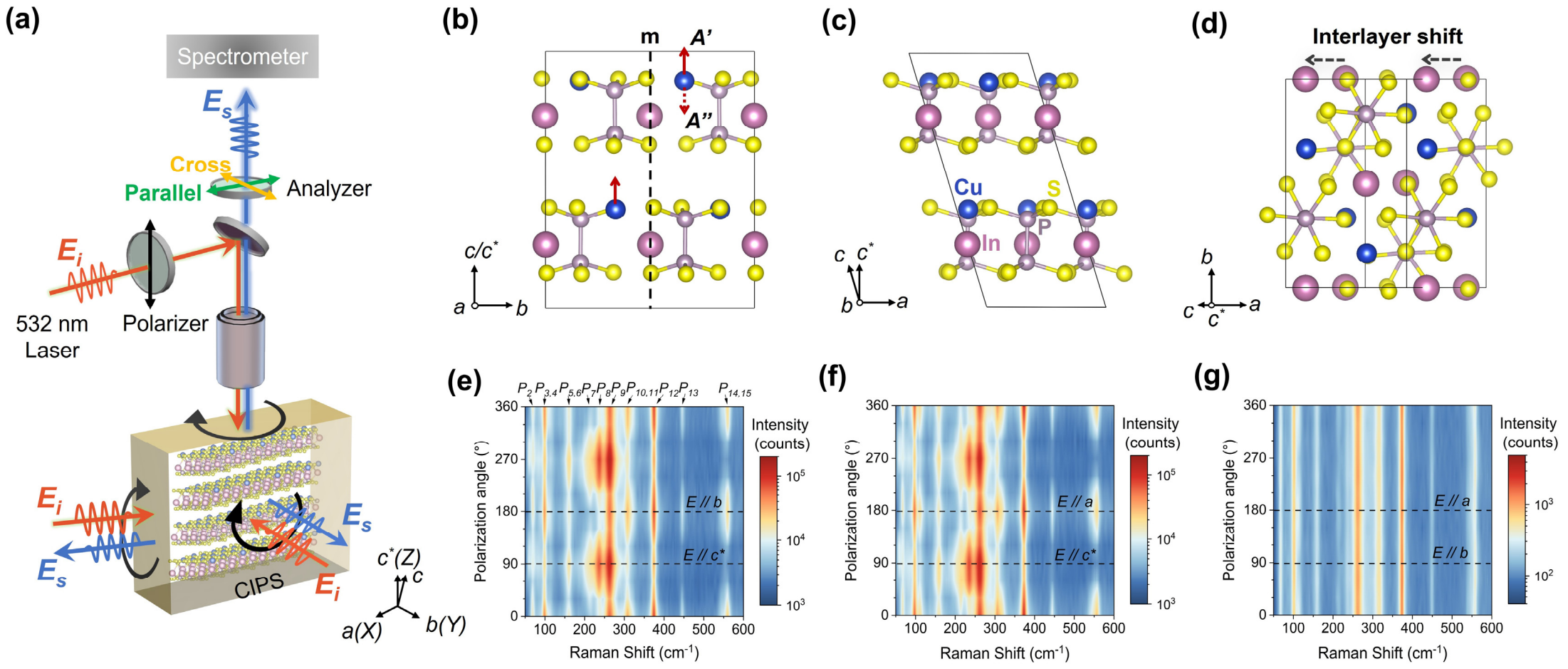

**Figure 1. Angle-resolved polarized Raman spectra of different crystal planes in CIPS.** (a) Schematic setup of the angle-resolved polarized Raman scattering. Atomic structure of CIPS viewed (b) along ***a*** (perpendicular to ***bc****$^{*}$* plane), (c) along ***b*** (perpendicular to ***ac****$^{*}$* plane), and (d) along ***c****$^{*}$* (perpendicular to ***ab*** plane) axes, respectively. Dash line and red arrows in (b) denote the glide plane (**m**) and vibrations with different symmetries (A' or A''). (e-g) Contour plots of angle-dependent polarized Raman spectra (parallel-polarized setup, 50 – 600 $cm^{-1}$) on (e) ***bc****$^{*}$* plane, (f) ***ac****$^{*}$* plane, and (g) ***ab*** plane, respectively. The plots are shown in logarithm scale to accentuate weak peaks. The dash line in each plot indicates two orthogonal in-plane axes.

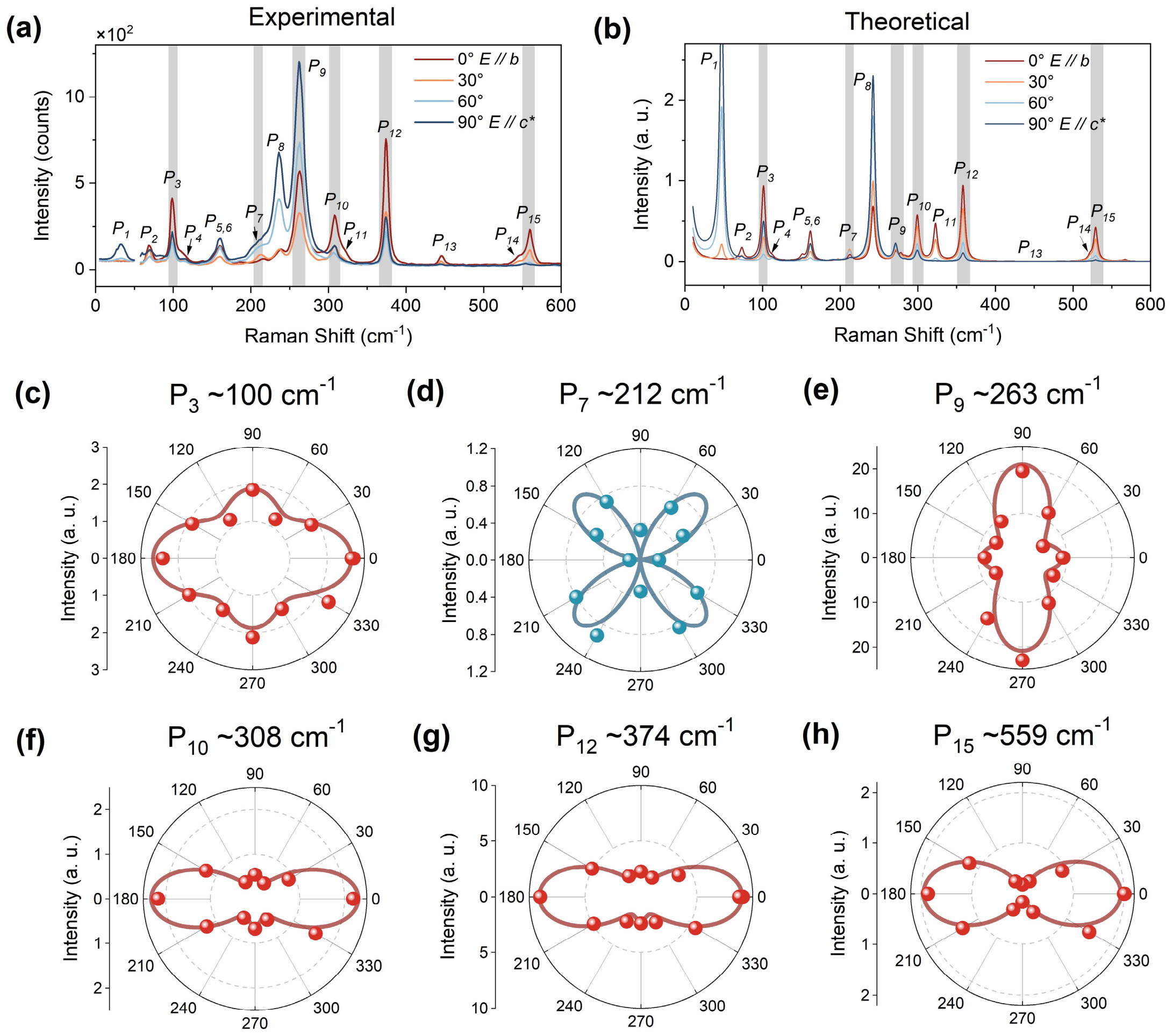

**Figure 2. Raman polarimetry of CIPS on $bc^*$ plane.** (a) Experimentally measured and (b) theoretically calculated polarized Raman spectra at selective angles. Main characteristic peaks $P_1 - P_{15}$ are labelled in the plot. (c-h) Polar plots of selective Raman peaks marked in grey in (a) and (b). Solid dots are experimental data and solid lines are corresponding fits according to the Raman tensor derived equations mentioned in the main text.

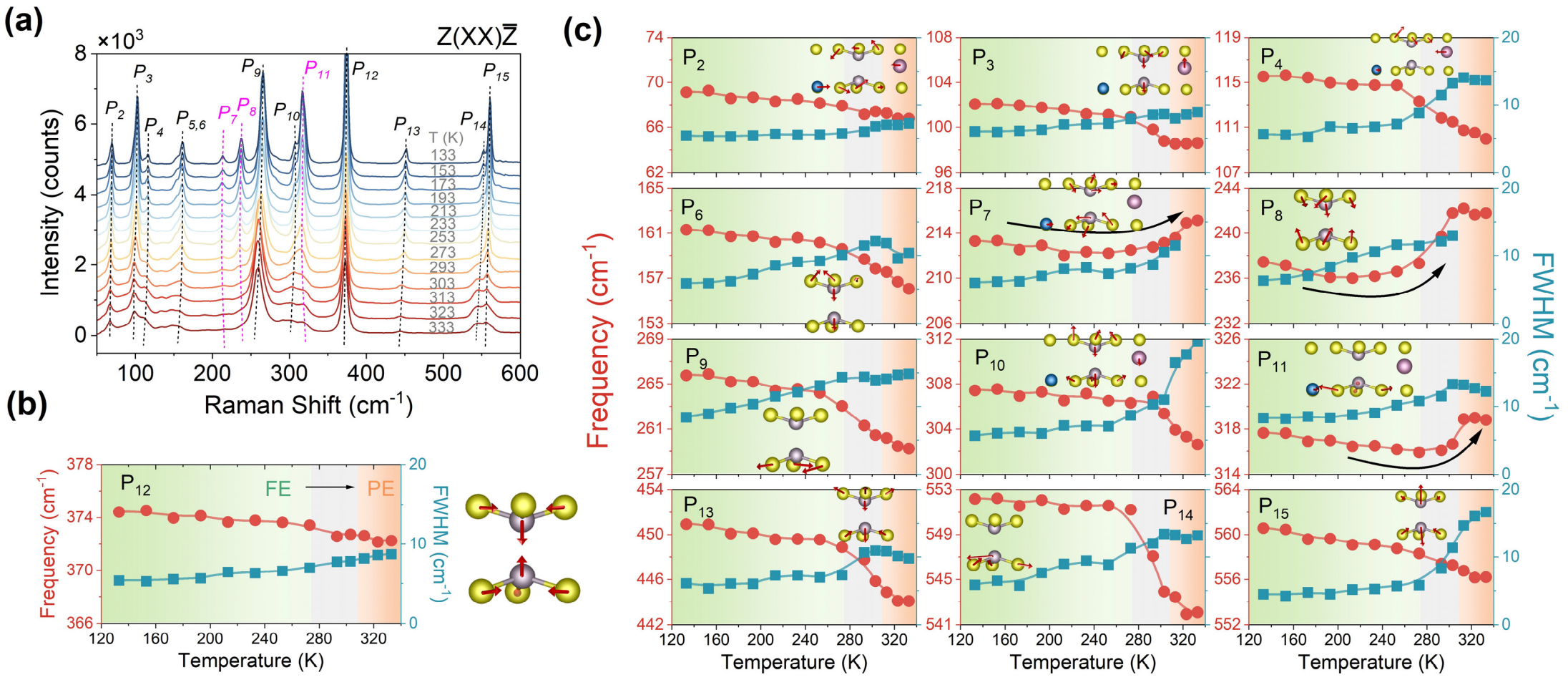

**Figure 3. Temperature dependent polarized Raman spectra of CIPS.** (a) Polarized Raman spectra of CIPS bulk crystal measured during temperature cooling from 333 to 133 K. The measurement was conducted in parallel-polarized setup with light incidence/scattering direction along $\boldsymbol{c}^*$ (Z) axis and its electric field of light along $\boldsymbol{a}$ (X) axis, that is, $Z(XX)\bar{Z}$ configuration. (b) Temperature dependent frequency and line width of Raman peak $P_{12}$ (P-P stretching mode) extracted from the spectra shown in (a). (c) Frequencies and line widths of other characteristic Raman peaks as a function of the temperature. All the plots have a fixed scale span at 12 $cm^{-1}$ for frequency (left axis) and 20 $cm^{-1}$ for FWHM (right axis), in order to facilitate the visualization of the temperature change rate for different peaks. The insets are cartoons depicting the displacement pattern of each mode.

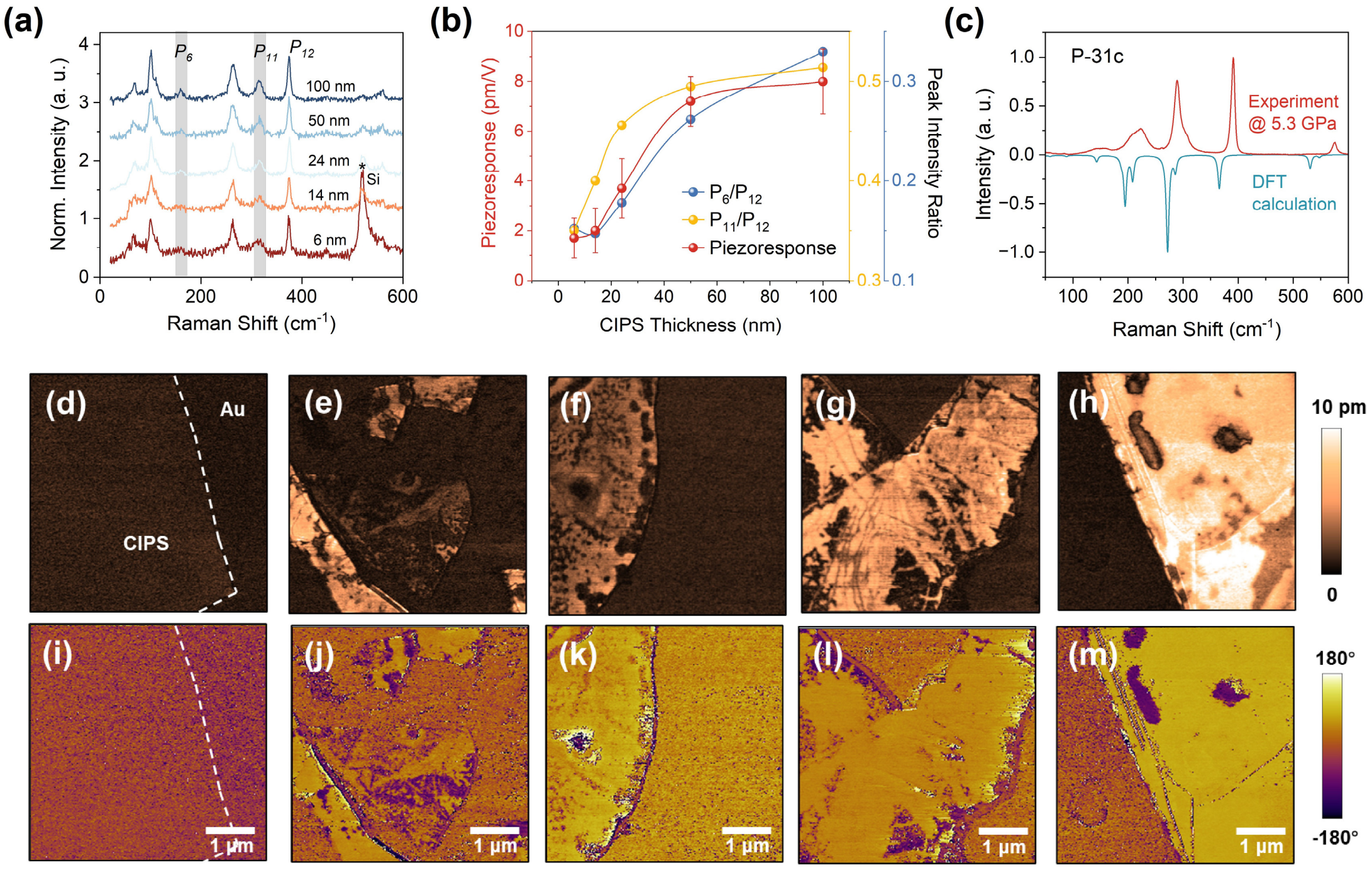

**Figure 4. Thickness dependent Raman scattering and PFM study of CIPS flakes.** (a) Raman spectra of CIPS flakes with different thicknesses measured under $Z(XX)\bar{Z}$ configuration at ambient condition. (b) A compilation plot of piezoresponse and Raman peak intensity ratios ($P_6/P_{12}$ and $P_{11}/P_{12}$) as a function of the flake thickness. (c) A comparison of experimental and calculated Raman spectra of trigonal $P\bar{3}1c$ phase of CIPS. Out-of-plane PFM (d-h) amplitude and (i-m) phase of CIPS flakes with varying thickness, (d, i) 6 nm, (e, j) 14 nm, (f, k) 24 nm, (g, l) 50 nm, and (h, m) 100 nm. The piezoresponse signal is quantitatively calibrated following the procedure described in Experimental Methods.

## Supplementary Materials

# Deciphering the lattice vibrational behaviors of $CuInP_2S_6$ by angle-resolved polarized Raman scattering

Yiqi Hu[1], Junjie Zhang[2,*], Zhou Zhou[1], Shun Wang[3], Qiankun Li[1], Yanfei Hou[1], Tianhao Ying[1], Lingling Yang[1], Jingyao Zhang[1], Shuzhong Yin[1], Yuyan Weng[1], Shuai Dong[2], Liang Fang[1,*], Lu You[1,3,*]

[1]School of Physical Science and Technology, Jiangsu Key Laboratory of Frontier Material Physics and Devices, Soochow University, Suzhou, 215006, China

[2]Key Laboratory of Quantum Materials and Devices of Ministry of Education, School of Physics, Southeast University, Nanjing 211189, China

[3]School of Electronic and Information Engineering, Suzhou University of Technology, Changshu 215500, PR China

*Corresponding authors: junjiezhang@seu.edu.cn; lfang@suda.edu.cn; lyou@suda.edu.cn

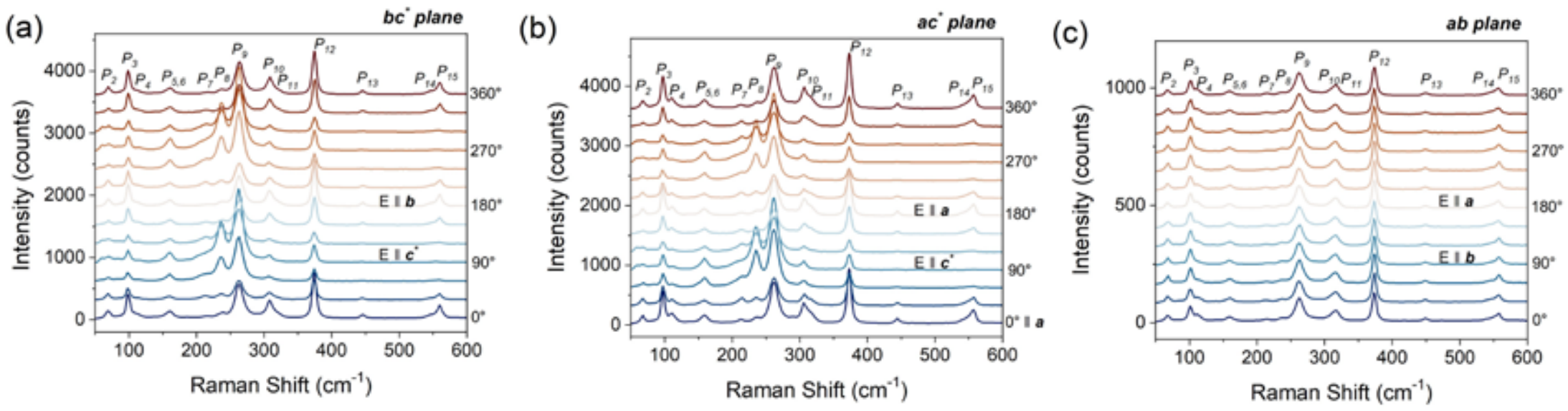

**Figure S1. Angle-dependent polarized Raman spectra measured on different crystallographic planes.** (a) ***bc****** plane, (b) ***ac****** plane, and ***ab*** plane.

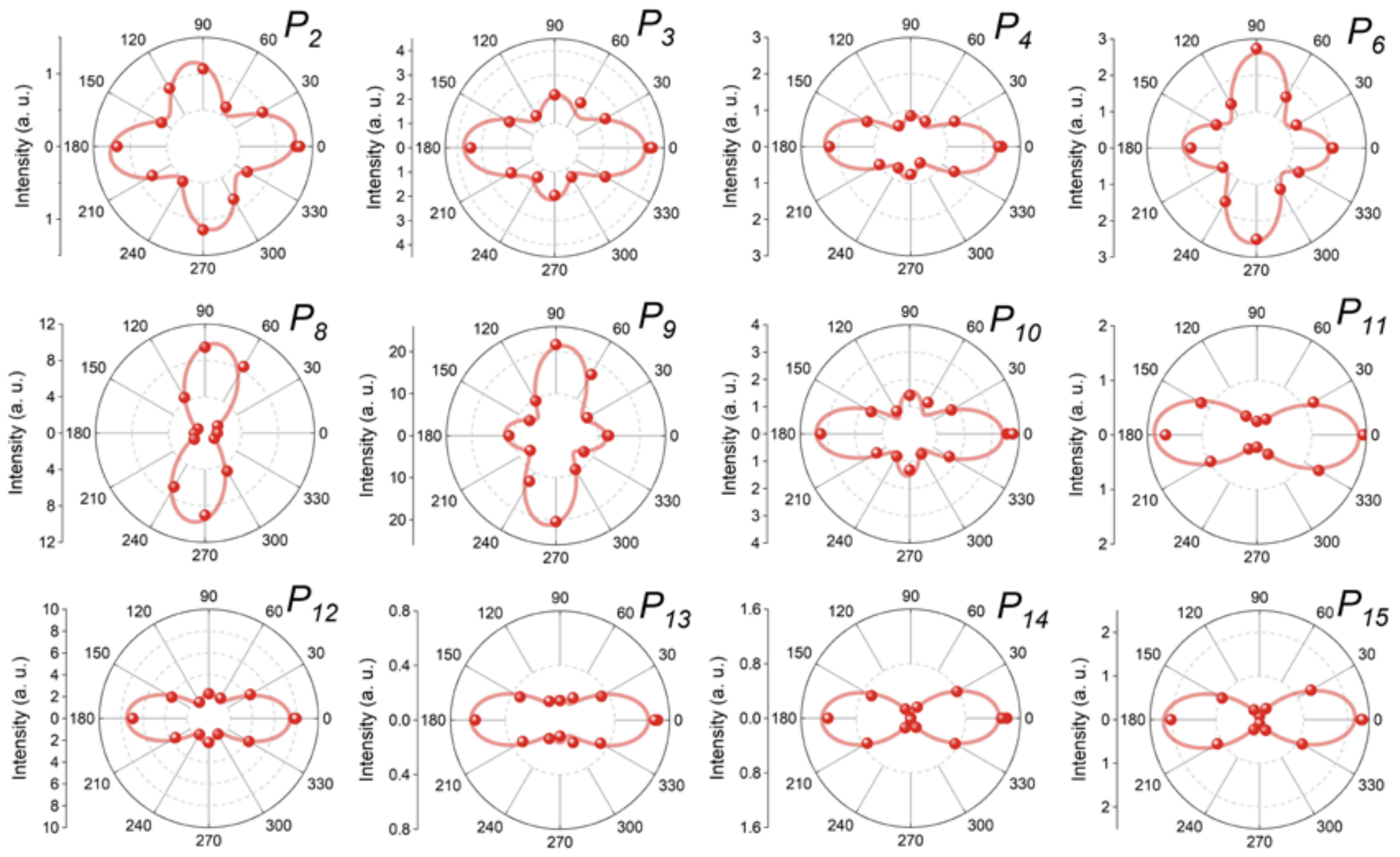

**Figure S2. Polar plots of main Raman peaks of CIPS measured on *ac** plane**. Solid dots are experimental data and solid lines are corresponding fits according to the Raman tensor derived equations.

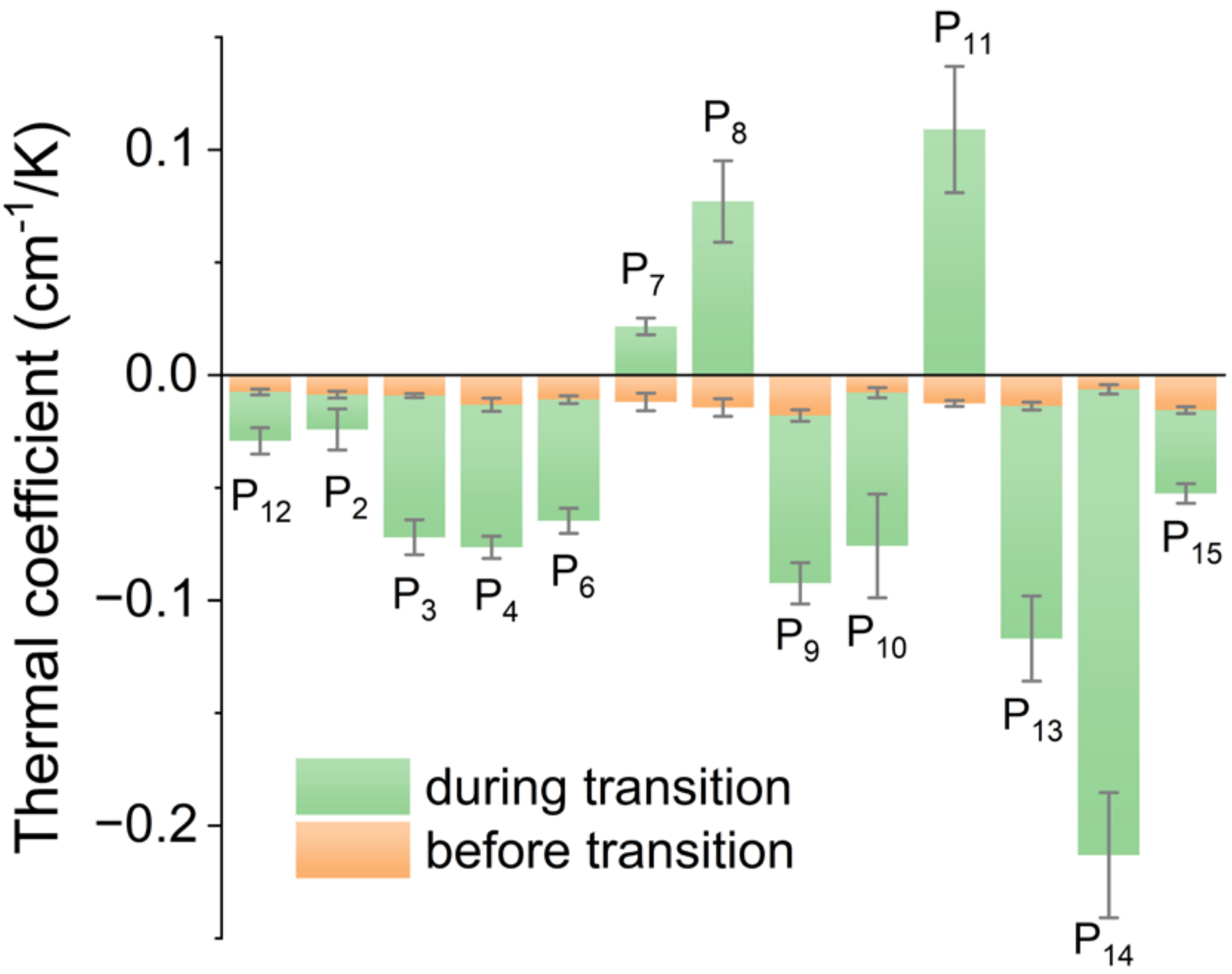

**Figure S3. Average thermal coefficients of different Raman peaks before and during FE-PE phase transition.**

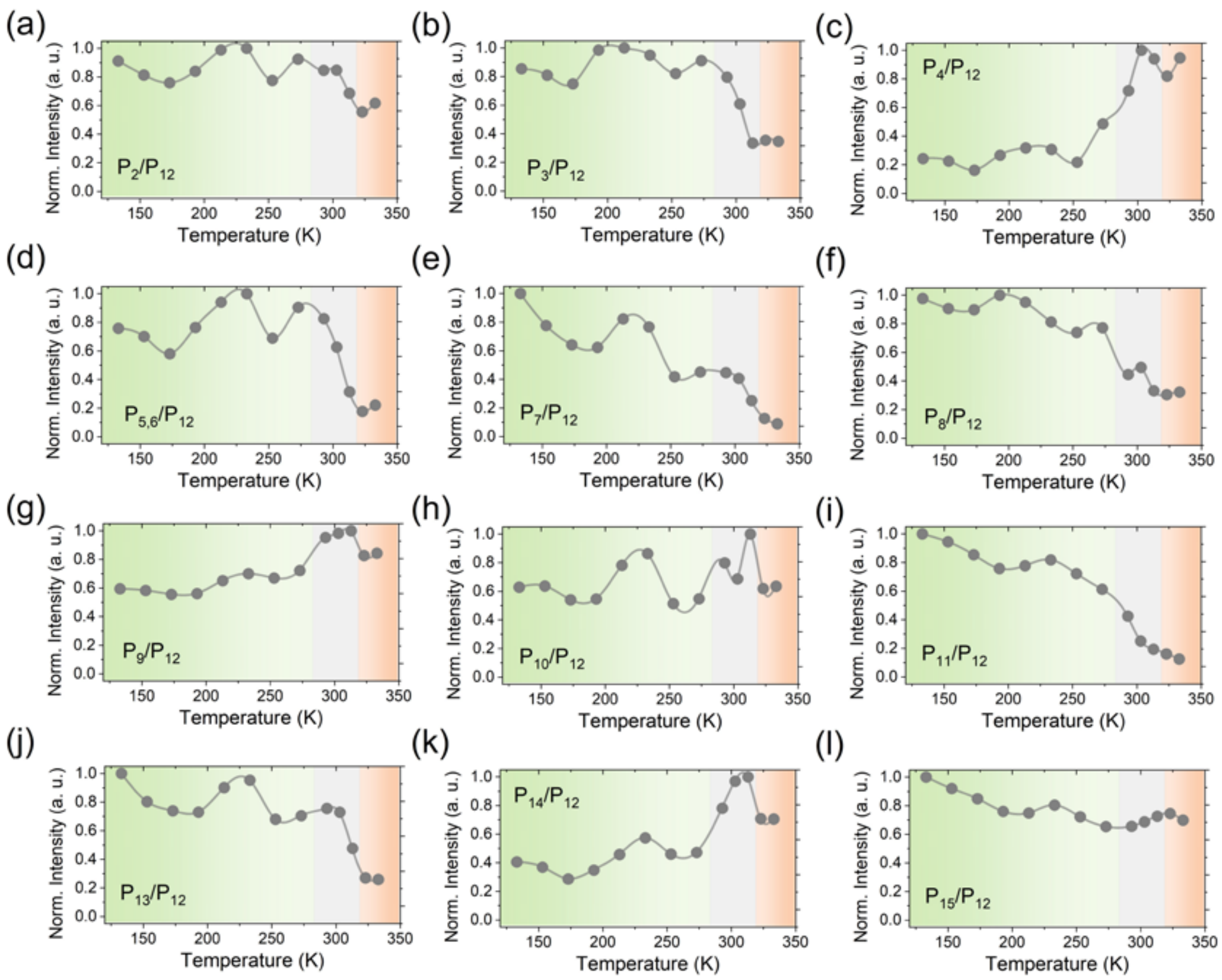

**Figure S4. Normalized intensity ratios of different Raman peaks with regard to $P_{12}$ as a function of temperature.** (a) $P_2$, (b) $P_3$, (c) $P_4$, (d) $P_{5,6}$, (e) $P_7$, (f) $P_8$, (g) $P_9$, (h) $P_{10}$, (i) $P_{11}$, (j) $P_{13}$, (k) $P_{14}$, (l) $P_{15}$.

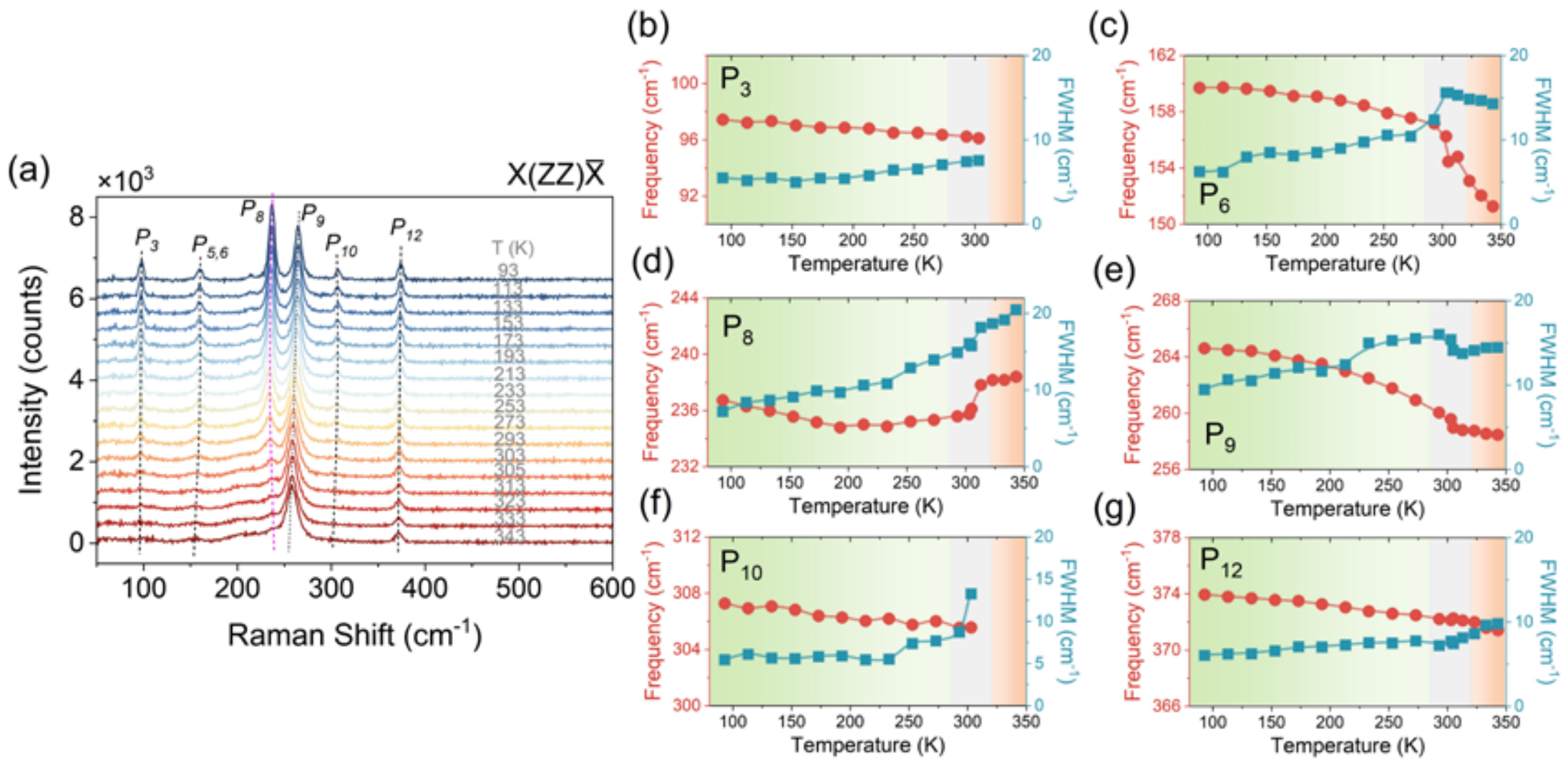

**Figure S5. Temperature dependent polarized Raman spectra of CIPS measured in $X(ZZ)\bar{X}$ configuration.** (a) Original Raman spectra recorded at various temperature. (b-g) Temperature dependent frequency and line width of characteristic Raman peaks, (b) $P_3$, (c) $P_6$, (d) $P_8$, (e) $P_9$, (f) $P_{10}$, and (g) $P_{12}$.

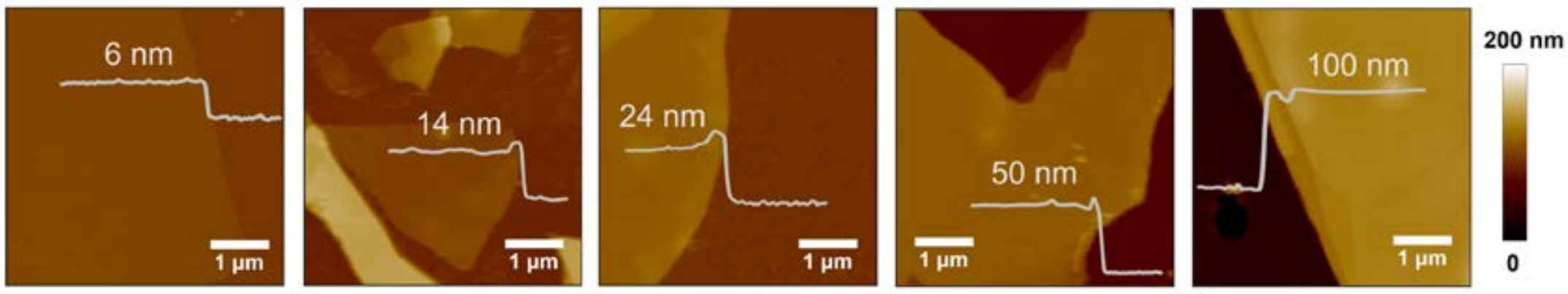

**Figure S6. Topographic images and thickness line profiles of CIPS thin flakes shown in Figure 4.**

**Table S1. DFT calculated Raman active phonon modes and their corresponding Raman tensors at 0 K.** The assigned Raman peaks discussed in the main text are highlighted in red. Corresponding experimental frequencies are given in parentheses.

| Phonon No. | Assigned Raman peak | Raman Shift ($cm^{-1}$) | Sym. | Raman tensor | | | | | |
|---|---|---|---|---|---|---|---|---|---|
| | | | | a | b | c | d | e | f |
| 4 | | 28.123 | A' | -0.02592 | -0.08413 | 0.09 | 0.01247 | -- | -- |
| 5 | | 28.58 | A'' | -- | -- | -- | -- | 0.09678 | -0.02426 |
| 6 | | 42.524 | A'' | -- | -- | -- | -- | 0.0041 | -0.12907 |
| 7 | $P_1$ | 47.356 (33) | A' | -0.24083 | 0.07663 | -5.08399 | -0.69344 | -- | -- |
| 8 | | 69.062 | A' | -0.59211 | -0.26032 | 0.27053 | -0.03369 | -- | -- |
| 9 | | 69.287 | A'' | -- | -- | -- | -- | 0.61921 | -0.00151 |
| 10 | | 70.708 | A'' | -- | -- | -- | -- | 0.91173 | 0.09433 |
| 11 | $P_2$ | 73.279 (69) | A' | 0.96116 | -1.57648 | 0.64617 | -0.21488 | -- | -- |
| 12 | | 73.689 | A'' | -- | -- | -- | -- | -0.63024 | -0.13712 |
| 13 | $P_3$ | 101.200 (99) | A' | 5.21062 | 5.34334 | -3.87209 | -0.08302 | -- | -- |
| 14 | | 109.834 | A'' | -- | -- | -- | -- | -0.3735 | -0.12758 |
| 15 | $P_4$ | 112.385 (112) | A' | 1.73721 | -1.02618 | -0.93173 | -0.98458 | -- | -- |
| 16 | | 113.215 | A'' | -- | -- | -- | -- | 0.47457 | -0.22062 |
| 17 | | 115.073 | A' | -1.24582 | -0.26249 | 0.19686 | -0.32194 | -- | -- |
| 18 | | 117.078 | A'' | -- | -- | -- | -- | 1.38184 | 0.76088 |
| 19 | | 150.399 | A'' | -- | -- | -- | -- | -0.81032 | 0.25619 |
| 20 | | 150.912 | A' | -0.32036 | 1.8501 | -1.15427 | -0.97472 | -- | -- |
| 21 | | 153.217 | A'' | -- | -- | -- | -- | -0.77578 | 0.52772 |
| 22 | | 153.672 | A' | -1.46915 | -0.10334 | 0.15747 | -0.38683 | -- | -- |
| 23 | | 160.215 | A'' | -- | -- | -- | -- | 0.50267 | -0.28371 |
| 24 | $P_6$ | 161.534 (160) | A' | -5.0415 | -5.08557 | 3.84499 | -0.15256 | -- | -- |
| 25 | | 189.777 | A'' | -- | -- | -- | -- | -0.48006 | -1.20858 |
| 26 | | 191.264 | A' | 0.28121 | -0.55031 | 0.73796 | 0.64276 | -- | -- |
| 27 | | 195.063 | A'' | -- | -- | -- | -- | -0.24567 | 0.44221 |
| 28 | | 196.899 | A' | -0.5361 | -1.30592 | -0.33285 | -1.31312 | -- | -- |
| 29 | | 211.224 | A' | -2.48985 | 1.55219 | -0.83366 | -1.91257 | -- | -- |
| 30 | $P_7$ | 211.553 (212) | A'' | -- | -- | -- | -- | 2.65688 | -3.62334 |
| 31 | | 212.875 | A' | 1.01629 | -2.47448 | -1.19482 | 3.42618 | -- | -- |
| 32 | | 212.919 | A'' | -- | -- | -- | -- | 1.34771 | 0.27484 |
| 33 | | 232.218 | A'' | -- | -- | -- | -- | 0.05386 | 0.37907 |
| 34 | $P_8$ | 242.325 (236) | A' | 9.19039 | 9.5237 | 17.60768 | -0.19327 | -- | -- |
| 35 | | 269.936 | A'' | -- | -- | -- | -- | -1.29428 | 0.46529 |
| 36 | $P_9$ | 271.369 (263) | A' | -3.38593 | -4.04844 | -5.71702 | 0.03786 | -- | -- |
| 37 | | 278.113 | A'' | -- | -- | -- | -- | 3.26262 | -1.2684 |
| 38 | | 278.154 | A' | 4.32621 | -3.30818 | 0.1205 | -1.98917 | -- | -- |
| **39** | | 283.937 | **A'** | -4.63435 | 2.12334 | -1.16753 | -0.50146 | -- | -- |
| **40** | | 284.553 | **A''** | -- | -- | -- | -- | 3.265 | 1.35652 |
| **41** | $P_{10}$ | 299.316 (308) | **A'** | -10.00409 | -10.20966 | -4.96447 | -0.26552 | -- | -- |
| **42** | | 309.384 | **A''** | -- | -- | -- | -- | -0.60598 | 0.13588 |
| **43** | | 322.118 | **A'** | 0.80064 | -0.64018 | 0.12394 | -1.6519 | -- | -- |
| **44** | | 322.368 | **A''** | -- | -- | -- | -- | -10.79708 | -0.93824 |
| **45** | $P_{11}$ | 322.777 (322) | **A'** | -10.57418 | 9.73858 | -0.7209 | 0.8683 | -- | -- |
| **46** | | 323.095 | **A''** | -- | -- | -- | -- | -0.16434 | 1.79137 |
| **47** | $P_{12}$ | 358.218 (374) | **A'** | -14.6497 | -15.0845 | -5.03126 | 0.05108 | -- | -- |
| **48** | | 358.642 | **A''** | -- | -- | -- | -- | -0.0873 | -0.11573 |
| **49** | $P_{13}$ | 430.851(446) | **A'** | -0.0151 | -0.02149 | -0.02672 | -0.00436 | -- | -- |
| **50** | | 438.692 | **A''** | -- | -- | -- | -- | -0.31518 | -0.45987 |
| **51** | $P_{14}$ | 521.981 (545) | **A'** | -4.70597 | -4.68141 | -1.22373 | 0.97746 | -- | -- |
| **52** | | 522.273 | **A''** | -- | -- | -- | -- | 2.54793 | -1.83236 |
| **53** | | 523.629 | **A'** | -5.61103 | 0.25967 | -0.2392 | -2.12797 | -- | -- |
| **54** | | 523.992 | **A''** | -- | -- | -- | -- | -0.7591 | 1.85711 |
| **55** | | 528.588 | **A''** | -- | -- | -- | -- | -0.07182 | 0.1621 |
| **56** | $P_{15}$ | 529.104 (559) | **A'** | 11.99121 | 13.16349 | 2.74078 | -0.09671 | -- | -- |
| **57** | | 560.534 | **A''** | -- | -- | -- | -- | -2.82334 | -1.59542 |
| **58** | | 560.542 | **A'** | -0.01286 | -1.80386 | 0.76715 | -0.39907 | -- | -- |
| **59** | | 567.19 | **A'** | 2.39788 | -3.01767 | 0.6461 | -1.40768 | -- | -- |
| **60** | | 571.34 | **A''** | -- | -- | -- | -- | 0 | 0 |

**Table S2. Assignments of experimental Raman peaks to DFT-calculated phonon modes with direct visualization of the displacement patterns and comparisons of Raman tensor ratios. (see attached word file)**

**Table S3. Temperature coefficient of each Raman peak extracted from Figure 3 before and during phase transition.**

| Raman peaks | T coefficient (before transition) ($cm^{-1}/K$) | T coefficient (during transition) ($cm^{-1}/K$) |
|---|---|---|
| $P_{12}$ | -0.008 ± 0.001 | -0.022 ± 0.006 |
| $P_2$ | -0.009 ± 0.002 | -0.016 ± 0.009 |
| $P_3$ | -0.009 ± 0.001 | -0.063 ± 0.008 |
| $P_4$ | -0.013 ± 0.003 | -0.063 ± 0.005 |
| $P_6$ | -0.011 ± 0.002 | -0.054 ± 0.006 |
| $P_7$ | -0.012 ± 0.004 | 0.022 ± 0.004 |
| $P_8$ | -0.014 ± 0.004 | 0.08 ± 0.02 |
| $P_9$ | -0.018 ± 0.003 | -0.074 ± 0.009 |
| $P_{10}$ | -0.008 ± 0.002 | -0.07 ± 0.02 |
| $P_{11}$ | -0.013 ± 0.001 | 0.11 ± 0.03 |
| $P_{13}$ | -0.014 ± 0.002 | -0.103 ± 0.02 |
| $P_{14}$ | -0.006 ± 0.002 | -0.21 ± 0.03 |
| $P_{15}$ | -0.016 ± 0.002 | -0.037 ± 0.004 |